\documentclass[11pt,letterpaper]{article}
\usepackage[margin=1in]{geometry}
\usepackage[T1]{fontenc}
\usepackage{amsmath,amssymb,graphicx,array,booktabs,longtable,tabularx}
\usepackage[hidelinks,hypertexnames=false]{hyperref}
\usepackage{caption,microtype}
\renewcommand{\arraystretch}{1.12}

\newcommand{\tableheading}[1]{\par\ifdim\dimexpr\pagegoal-\pagetotal\relax<120pt\newpage\fi\noindent#1\par\nopagebreak[4]}
\hypersetup{pdftitle={AIPOCH Open-Science: An Open-Source AI Research Workbench with Reproducible .science Packages},pdfauthor={AIPOCH Open-Science authors}}
\begin{document}
\begin{center}
{\LARGE AIPOCH Open-Science: An Open-Source AI Research Workbench with Reproducible .science Packages\par}
\vspace{18pt}
{\fontsize{10}{12}\selectfont Wen Ouyang${}^{1,17,\dagger}$, Bocheng Huang${}^{1,\dagger}$, Qianyu Yao${}^{1,\dagger}$, Fei Sun${}^{1,\dagger}$, Minying Deng${}^{2,\dagger}$, Haoyue Cheng${}^{2}$, Wenwen Zhou${}^{1}$, Lei Mou${}^{1}$, Yusi Wei${}^{1}$, Jiang Li${}^{1}$, Kun Zhu${}^{2}$, Lei Wang${}^{3,4,5}$, Yi Feng${}^{6}$, Jiaheng Dai${}^{7}$, Ke Qian${}^{8,5,9}$, Dan Wang${}^{10}$, Yahui Wang${}^{11}$, Chenyi Wang${}^{12}$, Mingsheng Lyu${}^{13}$, Yaqiang Zhang${}^{14}$, Ping Zheng${}^{45,15}$, Chunhua Tang${}^{16}$, Xinge Ma${}^{18}$, Xu Zhang${}^{19}$, Youjia Fan${}^{20}$, Can Lu${}^{21}$, Zhijun Dai${}^{22}$, Man Hu${}^{23}$, Qiqi Xie${}^{24}$, Beinan Liu${}^{70}$, Yuanning Guo${}^{67}$, Yunze Liu${}^{48}$, Jinhong Jiang${}^{69,68}$, Xianjun Fu${}^{35,36}$, Mingwei Zhang${}^{32}$, Yuanji Xu${}^{33}$, Shilin Xia${}^{56}$, Bowen Yang${}^{52,53,54}$, Mingjun Yu${}^{55}$, Jingdun Xie${}^{39}$, Xuwei Chen${}^{40}$, Xia Zou${}^{58}$, Yuzhou Xue${}^{59}$, Jianguo Zhou${}^{60}$, Chunhao Cao${}^{46,47}$, Donghao Tian${}^{61}$, Lei Huang${}^{49}$, Jinxiao Li${}^{37}$, Shengrong Long${}^{38}$, Zefeng Song${}^{57}$, Honglin Li${}^{50}$, Jiang Luo${}^{26}$, Wenying Jiang${}^{28}$, Shanshan Li${}^{27}$, Lei Dai${}^{29}$, Ning Jiang${}^{62}$, Ying Cai${}^{63}$, Chaohui Li${}^{41}$, Chunyan Peng${}^{42}$, Hetong Zhao${}^{25}$, Angqing Li${}^{31}$, Chunyan Yang${}^{51}$, Chongyang Sun${}^{64}$, Yong Li${}^{30}$, Xinyu Shi${}^{65}$, Wei Dai${}^{44}$, Ming Peng${}^{43}$, Chen Sang${}^{34}$, Yuanyuan Han${}^{66}$, Desong Qiu${}^{71}$, Haixin Miao${}^{71}$, Jian Xu${}^{72}$, Yi Chen${}^{73}$, Xiaoliang Shi${}^{74}$, Tian Shen${}^{75,76}$, Teng Hu${}^{77}$, Shuhao Wang${}^{77}$, Yingyong Hou${}^{2,*}$, Huimei Wang${}^{1,2,*}$\par}
\vspace{9pt}
{\small \textdagger{} These authors contributed equally to this work and share first authorship.\par
* Corresponding authors.\par}
\vspace{8pt}
{\small \textsuperscript{1} AIPOCH PTE. LTD., Singapore\par}
\vspace{3pt}
{\footnotesize Full author affiliations are listed on the following pages.\par}
\end{center}

\begin{abstract}
AI-assisted research leaves instructions, computations, artifact revisions and checks whose relationships matter when another researcher receives the work. We present AIPOCH Open-Science, an open-source research workbench with portable \texttt{.science} session packages, and specify an exchange contract for these recorded relationships. Version and assessment targets remain attributable while receiving-workspace identifiers can be remapped, eligible contents can be omitted explicitly, and imported history remains separate from a writable fork. A controlled session tested an incorrect first version, a reviewed correction, an unreviewed revision and a planned but unexecuted step. We compared complete and partial exports with source records extracted independently of the exporter, then inspected imports into separate projects within the same installation. Both conditions retained the tested version and review associations, kept the unreviewed revision distinct from earlier assessments and represented excluded content as unavailable. A documentation-rich ordinary directory also supplied evidence for all seven static inspection questions. A separate registry case connected a table to its recorded task, execution and review; offline execution of its retained script and fixed input reproduced two saved CSV tables byte for byte. The study contributes an implemented exchange contract and a controlled examination of its preservation rules, content-selection constraints and interpretation boundaries.
\end{abstract}
\clearpage
\section*{Author affiliations}
\begingroup
\small
\setlength{\tabcolsep}{3pt}
\renewcommand{\arraystretch}{1.12}
\begin{longtable}{@{}r p{0.92\linewidth}@{}}
\textsuperscript{1} & AIPOCH PTE. LTD., Singapore \\
\textsuperscript{2} & Department of Pathology, Zhongshan Hospital, Fudan University, Shanghai 200032, China \\
\textsuperscript{3} & Department of Breast Surgery, Fudan University Shanghai Cancer Center, Fudan University, Shanghai, China \\
\textsuperscript{4} & Key Laboratory of Breast Cancer in Shanghai, Fudan University Shanghai Cancer Center, Fudan University, Shanghai, China \\
\textsuperscript{5} & Department of Oncology, Shanghai Medical College, Fudan University, Shanghai, China \\
\textsuperscript{6} & Department of Integrative Medicine and Neurobiology, School of Basic Medical Sciences, Shanghai Medical College, Institutes of Integrative Medicine of Fudan University, Shanghai 200032, China \\
\textsuperscript{7} & College of Biomedical Engineering, Fudan University, Shanghai, China \\
\textsuperscript{8} & Department of Urology, Fudan University Shanghai Cancer Center, Fudan University, Shanghai 200032, China \\
\textsuperscript{9} & Shanghai Genitourinary Cancer Institute, Shanghai 200032, China \\
\textsuperscript{10} & Department of Pediatric Dentistry, Peking University School and Hospital of Stomatology, Beijing 100081, China \\
\textsuperscript{11} & Department of Physical Medicine and Rehabilitation, Beijing Tsinghua Changgung Hospital, School of Clinical Medicine, Tsinghua Medicine, Tsinghua University, Beijing 102218, China \\
\textsuperscript{12} & Department of Ultrasound, China-Japan Friendship Hospital, Beijing, China \\
\textsuperscript{13} & Department of Respiratory, The Third Affiliated Hospital, Beijing University of Chinese Medicine, Xiaoguan Street 51, Beijing, China \\
\textsuperscript{14} & Department of Traditional Medicine, Beijing Tongren Hospital, Capital Medical University, Beijing 100730, China \\
\textsuperscript{15} & Department of Neurosurgery, Shanghai Pudong New Area People's Hospital, Shanghai Jiao Tong University School of Medicine, Shanghai, China \\
\textsuperscript{16} & International Peace Maternity and Child Health Hospital, Shanghai Key Laboratory of Embryo Original Diseases, School of Medicine, Shanghai Jiao Tong University, Shanghai, China \\
\textsuperscript{17} & Shanghai Institute of Hematology, State Key Laboratory of Medical Genomics, National Research Center for Translational Medicine at Shanghai, Ruijin Hospital, Shanghai Jiao Tong University School of Medicine, Shanghai 200025, China \\
\textsuperscript{18} & Shanghai Advanced Institute of Finance, Shanghai Jiao Tong University, Shanghai, China \\
\textsuperscript{19} & Renji Hospital, Shanghai Jiao Tong University School of Medicine, Shanghai, China \\
\textsuperscript{20} & Department of Anesthesiology, Ruijin Hospital, Shanghai Jiao Tong University School of Medicine, Shanghai 200025, China \\
\textsuperscript{21} & Department of Colorectal Surgery and Oncology (Key Laboratory of Cancer Prevention and Intervention, China National Ministry of Education; Key Laboratory of Molecular Biology in Medical Sciences, Zhejiang Province, China), The Second Affiliated Hospital, Zhejiang University School of Medicine, Hangzhou, Zhejiang, China \\
\textsuperscript{22} & Department of Breast Surgery, The First Affiliated Hospital, College of Medicine, Zhejiang University, Hangzhou 310000, China \\
\textsuperscript{23} & Department of Internal and Vascular Medicine, Amsterdam UMC, University of Amsterdam, Amsterdam 1105 AZ, The Netherlands \\
\textsuperscript{24} & School of Information Sciences, University of Illinois Urbana-Champaign, Champaign, IL 61820, USA \\
\textsuperscript{25} & Department of Traditional Chinese Medicine, No. 905 Hospital of Chinese People's Liberation Army Navy, Naval Military Medical University, Shanghai, China \\
\textsuperscript{26} & East China University of Science and Technology, Shanghai, China \\
\textsuperscript{27} & Department of Internal Medicine of Traditional Chinese Medicine, The First Clinical Medical College, Nanjing University of Chinese Medicine, Nanjing, China 210023 \\
\textsuperscript{28} & Department of Breast Surgery, The Third Affiliated Hospital of Soochow University, Changzhou, China \\
\textsuperscript{29} & Department of Hepato-pancreato-biliary Surgery, Ningbo Medical Centre Lihuili Hospital, The Affiliated Hospital of Ningbo University, Ningbo 315040, Zhejiang, China \\
\textsuperscript{30} & Department of Clinical Laboratory, First Affiliated Hospital of Huzhou Normal University, The First People's Hospital of Huzhou City, Huzhou, Zhejiang, China \\
\textsuperscript{31} & Department of General Surgery, The Third Affiliated Hospital of Anhui Medical University (The First People's Hospital of Hefei), Hefei, Anhui 230061, China \\
\textsuperscript{32} & Department of Radiotherapy, The First Affiliated Hospital of Fujian Medical University, Fuzhou 350005, Fujian, China \\
\textsuperscript{33} & Department of Radiation Oncology, Clinical Oncology School of Fujian Medical University, Fujian Cancer Hospital, Fuzhou 350014, China \\
\textsuperscript{34} & Department of Head and Neck Surgery, Key Laboratory of Head \& Neck Cancer Translation Research of Zhejiang Province, Zhejiang Cancer Hospital, Hangzhou Institute of Medicine (HIM), Chinese Academy of Sciences, Hangzhou, Zhejiang 310022, China \\
\textsuperscript{35} & Marine Traditional Chinese Medicine Research Center, Qingdao Academy, Shandong University of Traditional Chinese Medicine, Qingdao 266114, China \\
\textsuperscript{36} & Shandong Engineering and Technology Research Center of Traditional Chinese Medicine, Jinan 250355, China \\
\textsuperscript{37} & Union Hospital, Tongji Medical College, Huazhong University of Science and Technology, Wuhan 430022, Hubei, China \\
\textsuperscript{38} & Department of Thoracic Surgery, Tongji Hospital, Tongji Medical College, Huazhong University of Science and Technology, Wuhan, China \\
\textsuperscript{39} & Department of Anesthesiology, Sun Yat-sen University Cancer Center, Guangzhou 510060, China \\
\textsuperscript{40} & Department of Breast and Thyroid Surgery, The Third Affiliated Hospital of Sun Yat-sen University, Guangzhou 510700, China \\
\textsuperscript{41} & The Third Affiliated Hospital, Guangzhou Medical University, Guangzhou, China \\
\textsuperscript{42} & Clinical Molecular Diagnostic Center of Taihe Hospital, Hubei University of Medicine, China \\
\textsuperscript{43} & Wuhan Fourth Hospital, Wuhan, China \\
\textsuperscript{44} & Shangrao Key Laboratory of Acute and Critical Care Medicine, Jiangxi Medical College, Shangrao, China \\
\textsuperscript{45} & CHMD, Hong Kong Baptist University, Hong Kong SAR, China \\
\textsuperscript{46} & Institute of Integrated Bioinfomedicine and Translational Science (IBTS), School of Chinese Medicine, Hong Kong Baptist University, Hong Kong SAR, China \\
\textsuperscript{47} & Department of Systems Biology, School of Life Sciences, Southern University of Science and Technology, Shenzhen 518055, China \\
\textsuperscript{48} & Department of Biomedical Science, Faculty of Medicine, University of Macau, Macau SAR, China \\
\textsuperscript{49} & Department of Cardiology, The Third Central Hospital of Tianjin; Tianjin University Central Hospital; Tianjin Key Laboratory of Extracorporeal Life Support for Critical Diseases; Artificial Cell Engineering Technology Research Center; Tianjin Institute of Hepatobiliary Disease; Nankai University Affiliated Third Central Hospital, Tianjin, China \\
\textsuperscript{50} & The Fifth Affiliated Hospital of Zhengzhou University, China \\
\textsuperscript{51} & Department of Obstetrics and Gynecology, Xingtai People's Hospital, Affiliated to Hebei Medical University, Xingtai, Hebei, China \\
\textsuperscript{52} & Department of Medical Oncology, First Hospital of China Medical University, Shenyang, China \\
\textsuperscript{53} & Key Laboratory of Anticancer Drugs and Biotherapy of Liaoning Province, First Hospital of China Medical University, Shenyang, China \\
\textsuperscript{54} & National Institute of Health and Medical Big Data of China Medical University, Shenyang, China \\
\textsuperscript{55} & Department of Neurosurgery, Shengjing Hospital of China Medical University, Shenyang 110004, China \\
\textsuperscript{56} & Department of General Surgery, Clinical Laboratory of Integrative Medicine, The First Affiliated Hospital of Dalian Medical University, Dalian, China \\
\textsuperscript{57} & Faculty of Medicine, Dalian University of Technology, Dalian 116081, China \\
\textsuperscript{58} & Department of Science and Technology, West China Hospital, Sichuan University, Chengdu 610041, China \\
\textsuperscript{59} & Department of Cardiology, West China Hospital, Sichuan University, Chengdu, China \\
\textsuperscript{60} & Department of Oncology, The Second Affiliated Hospital of Zunyi Medical University, Zunyi, China \\
\textsuperscript{61} & Department of Pediatric Surgery, The Second Affiliated Hospital of Xi'an Jiao Tong University, Xi'an 710004, Shaanxi, China \\
\textsuperscript{62} & Department of Pathology, College of Basic Medical Sciences, Chongqing Medical University, Chongqing 400016, China \\
\textsuperscript{63} & School of Nursing, Chongqing Medical University, Chongqing 400016, China \\
\textsuperscript{64} & The First Affiliated Hospital of Harbin Medical University, China \\
\textsuperscript{65} & Department of Medical Science of Laboratory, Liaoning University of Traditional Chinese Medicine, Shenyang, China \\
\textsuperscript{66} & Institute of Medical Biology, Chinese Academy of Medical Sciences and Peking Union Medical College, Kunming, Yunnan 650118, China \\
\textsuperscript{67} & Department of Genetics and Developmental Biology, The Ruth and Bruce Rappaport Faculty of Medicine, Technion -- Israel Institute of Technology, Haifa 31096, Israel \\
\textsuperscript{68} & Department of Neurology, Beth Israel Deaconess Medical Center, Boston, MA 02215, USA \\
\textsuperscript{69} & Jiangsu Province Key Laboratory of Anesthesiology, Xuzhou Medical University, 209 Tongshan Rd, Xuzhou, Jiangsu 221004, China \\
\textsuperscript{70} & International Medical Service Center (UK) \\
\textsuperscript{71} & Sichuan Smart Medicine Technology Co., Ltd., Chengdu, Sichuan 610041, China \\
\textsuperscript{72} & Shuzu Technology (Nanjing) Co., Ltd., Nanjing, Jiangsu, China \\
\textsuperscript{73} & Huayi Vocational Skills Training Center, Putuo District, Shanghai, China \\
\textsuperscript{74} & Amoy Diagnostics Co., Ltd, Xiamen 361000, China \\
\textsuperscript{75} & SenseTime Research, Shanghai, China \\
\textsuperscript{76} & SenseCare Research, Shanghai, China \\
\textsuperscript{77} & Thorough Lab, Thorough Future, Beijing, China \\
\end{longtable}
\addtocounter{table}{-1}
\endgroup

\clearpage
\begin{center}
\includegraphics[width=\linewidth]{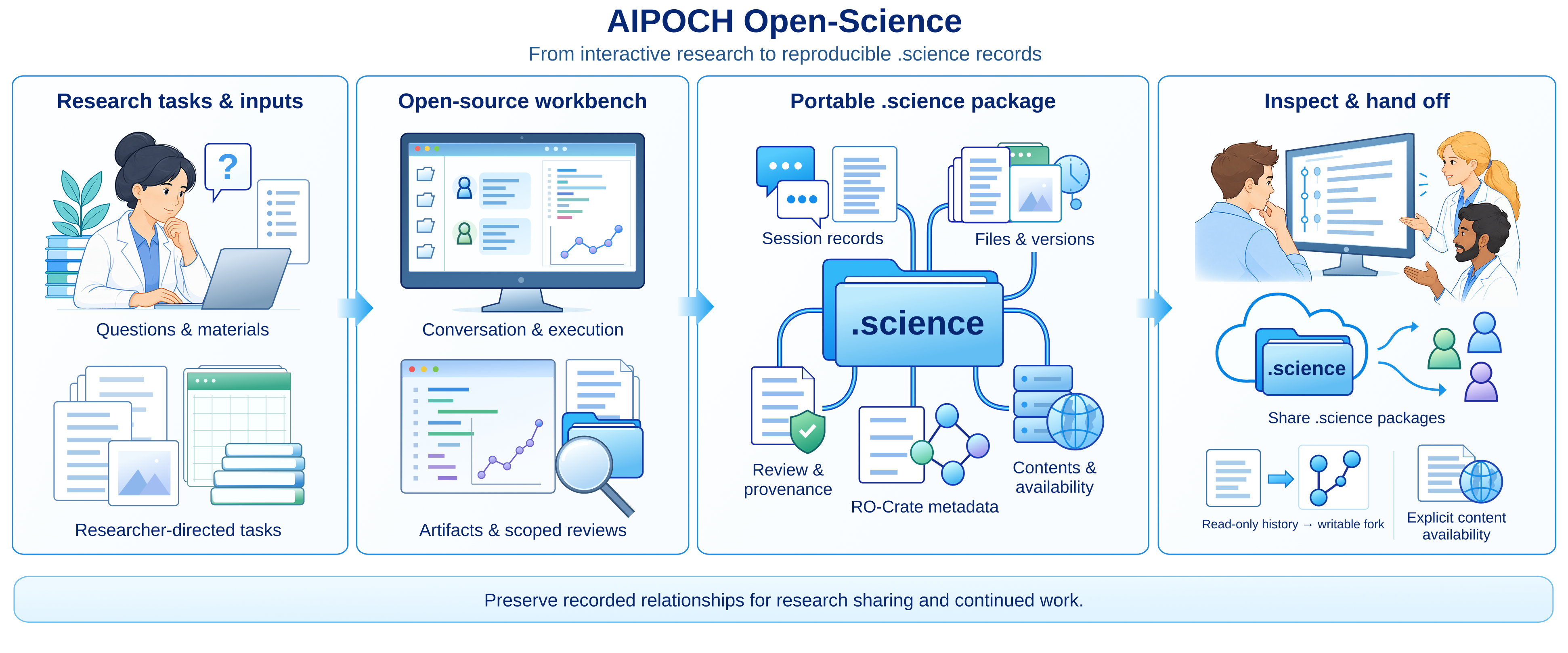}
\end{center}
\noindent\textbf{Graphical abstract.} Research tasks, persistent workbench records and reproducible \texttt{.science} packages for inspection and handoff.
\bigskip
\section{Introduction}
An AI-assisted scientific result is produced through a sequence of research instructions, computations and decisions. A researcher may specify a task, approve a plan, inspect an intermediate table and request a correction before accepting a report. The agent's code and tools execute some of these instructions; the resulting files may then be revised and checked. A collaborator receiving the output needs to establish which task it answers, which recorded activity produced it, which version was checked and which underlying materials remain available. These questions make the research session an important object of sharing.

Mainstream coding agents already provide substantive computational assistance. Studies of Claude Code and Codex found that the agents could reproduce many social-science findings when adequate materials were supplied [1]. Material availability and prompt framing affected interpretation of those outcomes. A separate many-analysts study distinguished methodological choices from the rules used to turn estimates into substantive conclusions [2]. Changed verdicts did not necessarily accompany comparable changes in numerical estimates. These findings motivate retaining the context in which scientific outputs are produced and interpreted. A researcher can assign a scientific task to a capable coding agent; surrounding infrastructure must organize the evidence left by that work.

Computational reproducibility depends on data, code and runtime conditions. A recent review distinguishes LLMs that generate durable artifacts from LLMs that remain live dependencies of a published calculation [3]. Executable documents preserve code and output together, but their presence alone does not establish rerunnability. A large study of biomedical Jupyter notebooks found frequent failures associated with dependencies and execution conditions [4]. Research exchange therefore requires identifying available objects and retaining the relationships needed to interpret and reuse them.

Provenance and research-object infrastructure already addresses much of this need. W3C PROV describes entities, activities and responsible agents, while P-Plan connects intended plans to recorded executions [5,6]. RO-Crate packages research artifacts with contextual metadata, and Workflow Run RO-Crate describes computational runs [7,8]. AiiDA maintains provenance graphs and applies consistency rules when exporting selected nodes [9,10]. More recent agent systems record prompts, actions, artifacts and evaluation context [11--15]. These foundations motivate a specific exchange problem. Recorded relationships must retain their meaning across artifact revisions, unavailable materials and transfer of selected contents to another workspace.

Three conflicts define this question. First, a planned action may lack an execution record; the plan's presence must not be read as evidence that its calculation occurred. Second, a finding about an earlier artifact version remains a valid historical record, but does not automatically assess a later revision. Third, selective file sharing may preserve a record's identity while removing access to its payload. A package must make this distinction visible or reject a selection that contradicts the evidence it retains. These conflicts concern the relation between recording, sharing and interpreting evidence, rather than the task score of the model that generated it.

AIPOCH Open-Science and \texttt{.science} address these conflicts through persistent workbench records and an importable session package. The contribution has two parts. First, we specify and implement an exchange contract that distinguishes relations that must remain attributable from workspace-dependent properties that may change. Version identity in a lineage, revision direction and assessment targets remain distinguishable across local identifier mapping; eligible payload omissions remain explicit, and imported history has a separate writable continuation context. Relationship-dependent capture and content-selection guards implement these rules.

Second, we examine the contract through a controlled trajectory containing an incorrect output, a reviewed correction, an unreviewed revision, an unexecuted plan and selective sharing. Paired source, export and receiver records test the specified associations, while application observations examine version-specific checks and the history/fork boundary. The resulting observations identify both preserved relations and limits: initial task attribution still needs surrounding conversation, observed file use does not establish a complete native dependency graph, and a required-evidence control prevents some content exclusions. A registry case supplies a separate inspection and fixed-input re-execution example. These findings give concrete design rules and observations for exchanging interactive research records.

\section{Related work}
\subsection{Scientific execution with coding agents and research workbenches}
Claude Code, Codex and related coding harnesses can manipulate repositories and execute analysis programs. Their scientific use makes task specifications, input snapshots and interpretation rules relevant parts of the retained record [1--3]. Biomedical and scientific agents add domain resources and workflow organization: Biomni integrates biomedical tools with code execution, while K-Dense Analyst describes planning and implementation loops with dedicated review [16--18]. Existing delivery capabilities include saved execution history and downloadable research materials [17]. The exchange problem considered here concerns the attribution and interpretation of retained records across revisions and content selection.

OpenAI4S is a close implemented precedent: it combines persistent scientific execution with versioned artifacts, environment records and session exchange, including quarantined read-only import [11]. Our design builds on this established session-oriented approach and examines the exchange of workbench-managed artifact versions and scoped reviews when selected contents differ from the retained record set. The evaluation tests this contract within AIPOCH Open-Science; it does not compare the two systems experimentally.

\subsection{Provenance representations and partial disclosure}
PROV and P-Plan establish the distinction between a plan and a past execution [5,6]. AiiDA provides a further precedent for export consistency: traversal rules include related nodes needed to maintain provenance [9,10]. Provenance abstraction research also studies how information can be hidden while preserving valid relationships and avoiding unjustified dependencies [19]. Partial sharing and graph consistency are therefore established research problems. Our implementation makes an operational choice for interactive sessions: selected native records and their required evidence remain together, while eligible immutable payloads can be omitted with explicit descriptors. Some selections are refused when the omitted bytes also occur in evidence that cannot safely be separated.

RO-Crate supplies a metadata packaging foundation [7]. Workflow Run RO-Crate profiles support execution descriptions at multiple granularities [8]; the Five Safes extension additionally records assessment and disclosure-control activities [20]. \texttt{.science} uses RO-Crate metadata alongside an application-native representation. Native records carry the structures needed by the receiving workbench; the standardized graph describes retained objects and recorded provenance. The two representations have complementary roles rather than identical detail. Certification against Workflow Run profiles was outside the present evaluation.

\subsection{Agent records, artifacts and evaluation evidence}
PROV-AGENT models agent interactions as part of workflow provenance, and AgentTrails derives artifact relationships from agent trajectories [12,13]. Artifact-centered, claim-aware observability distinguishes derivation, evidence, evaluation and steering, and proposes preservation of event identity when payload access is redacted [14]. F(AI)2R records AI and human contributions together with explicit verification states [15]. These precedents establish the importance of artifact identifiers and claim--evidence bindings. Our contribution concerns their implementation and validation at the boundary between a persisted research session and a selected, importable package.

Agent-Native Research Artifacts provide a broader publication-oriented precedent for organizing scientific logic, code, exploration and evidence [21]. The \texttt{.science} object instead retains a workbench session without requiring its conversion into a complete publication argument. Its inspection value depends on locatable relationships, while computational reuse depends on retained inputs and runtime conditions.

\section{Methods and implementation}
\subsection{Research setting and evaluation questions}
The study concerns researcher-directed tasks executed within an AI research workbench. Inputs comprise a task description and available research materials. Researchers and agents may add messages, plans, tool activities, Notebook computations, uploaded inputs, artifact versions and review findings. The exporter receives a session and a permitted selection of branches and contents. A compatible receiver imports preserved history and starts subsequent work in a separate writable context.

The evaluation distinguishes the intended event, captured source record, selected export record and receiving-workspace representation. Two research questions guide the design. First, how can the workbench retain task context, executed activities, artifact revisions and checks needed to inspect a researcher-directed process? Second, how can these relationships remain correctly attributable after exchange when artifacts change or selected contents are withheld? Evaluation requirements specify observable relationships and interpretation rules rather than a count of exported files. Source capture, exchange preservation and computational re-execution are assessed separately.

For evaluation, a session is represented as typed objects and relations mapped to actual native fields. Objects comprise task and plan records, execution activities, input and artifact versions, review scopes and findings. Relations describe task correspondence, execution status, input association, version production, revision direction and assessment scope. Association strength remains explicit. Attachment to a turn, observed resolver access and evidence of actual computational use describe different levels of support.

\subsection{The source workbench and why it maintains these records}
AIPOCH Open-Science separates the research record from the agent runtime that performs a task. Projects and sessions persist locally; compatible frameworks and providers supply configured agent execution. Managed Python and R computation, scientific connectors and task-relevant guidance support the work. The Literature Library organizes references, PDFs and notes that may be supplied as research inputs. These capabilities explain where session records originate, while the exchange method remains centered on the records selected for export.

The source workbench registers research files as artifact versions rather than identifying them solely by their current filename. Version records can refer to an earlier version, the task message, the producing activity or Notebook run, retained message snapshots and available execution evidence. Input associations retain their reported strength. The inspected implementation distinguishes turn-attached inputs from resolver-accessed inputs and records whether a producer association was agent-declared and validated or inferred from application observations. This qualification matters: a generic edge to an input must not silently upgrade an attachment into a proven numerical dependency.

Review records add a different relationship. A Reviewer finding can name an artifact version, while its parent review retains the assessed scope. Review-scope snapshots preserve the evidence presented during that assessment. Subsequent dispositions can connect an earlier finding to a later corrective review and the version that the correction assessed. The purpose is to preserve the history of checks without allowing a later workflow-level outcome to erase the assessment of an earlier file version.

The current review projection selects direct assessments of the requested version and separates them from later reviews in the linked history. Scope recomputation also distinguishes changed evidence from unavailable verification: failure to recompute a scope reports unavailable verification rather than establishing that the contents changed. These mechanisms support precise interpretation of a check. They do not turn the reviewer's judgment into an independently established scientific result.

\subsection{Relationship-dependent record capture}
The exporter uses an explicit native-table allowlist. It captures artifact lineages and versions, uploaded file versions, input associations, message snapshots, reviews, findings, finding dispositions and review-scope snapshots, together with supported computational and reproducibility records. The allowlist makes the export contract inspectable and avoids treating the entire local database as a portable research object.

Starting with selected session records and referenced versions, capture traverses recorded relationships until no new required objects are found. The inspected traversal follows earlier versions, upstream input versions, findings associated with selected artifacts, the reviews that own those findings and the reviews or artifact versions referenced by subsequent dispositions. It also collects versions explicitly named by a review scope. It resolves these exact upstream objects without automatically copying unrelated foreign conversations. Missing required source evidence causes the operation to fail.

This traversal addresses a practical sharing problem: transferring a finding without its target version or scope would make the check difficult to interpret. The traversal is consistent with prior provenance export principles [10]. Its application here includes the workbench's versioned review and conversation records, whose exact fields and capture rules can be examined in the implementation. The resulting set is bounded by recorded relations; a dependency never captured by the application cannot be supplied by traversal alone.

\subsection{Selection of contents and explicit availability}
The package separates record retention from retention of eligible file payloads. Metadata and integrity information remain required; immutable artifact, upload or supported reproducibility payloads can be selected for exclusion. The selection preview identifies related input users, descendants, reviews and Notebook runs so that a researcher can inspect the potential consequences of sharing a subset of contents.

An exclusion must name an eligible known storage key and match its recorded filename and size. Excluded content is listed explicitly in the manifest. This records a known identity whose bytes were intentionally withheld rather than presenting an unexplained gap. Missing materials have a separate omission description where supported, including unavailable environment-lock contents or inaccessible protected computational evidence. These descriptions provide boundaries for interpretation, without claiming to enumerate all dependencies that the source application never observed.

Some payloads occur more than once, for example as a file version and as retained Notebook or execution evidence. Excluding the visible artifact while retaining the same bytes elsewhere would contradict the selection. The implementation checks retained inventory digests and rejects conflicting exclusions where the evidence reader does not support a safe omission. Certain disposable Notebook input copies can be omitted with their source payload under a validated rule. This is a deliberate tradeoff: the system may refuse a smaller package to preserve the consistency of its retained evidence. The check detects matching retained bytes; it is not a general guarantee against every transformed or partial disclosure of sensitive information.

\subsection{Package representation and RO-Crate mapping}
In the inspected implementation, a \texttt{.science} file is a gzip-compressed tar archive. Its principal entries are \texttt{manifest.json}, \texttt{session.json}, \texttt{records.json}, \texttt{ro-crate-metadata.json} and a package-format \texttt{README.md}; retained objects are stored under \texttt{objects/}. The manifest identifies the native session format, schema version, source session, inventory and declared omissions. Checksums support integrity checks on retained contents.

The native session and record files retain the application structures needed for compatible historical import. RO-Crate metadata describes retained entities, files and recorded activities using JSON-LD. A retained file can be linked to its native artifact-version identity, allowing the reader to connect a named output, its bytes and its record. The metadata graph complements the native records; it does not replace them as the source for workbench import or create dependencies unsupported by the captured evidence.

The package provides format and schema identity, session and conversation records, artifact versions, recorded provenance relations, available execution and review evidence, and retained contents. These components answer different inspection questions. Conversation supplies task context, and execution records document activity. Version records distinguish revisions, while review records delimit the assessed scope. Content descriptors identify what a recipient can access and which contents were excluded. Their relationships, rather than a filename list alone, form the basis of the exchange method.

Two README roles support interpretation without changing the native exchange representation. The package-format README provides basic reading guidance. Hugging Face case \texttt{README.md} files are separately prepared explanations of public examples and do not participate in the export workflow.

\subsection{Export consistency and receiving-workspace control}
Export captures records, copies retained contents and generates their inventory. The implementation rechecks relevant source state before publication and refuses an export when the session or required evidence changed during capture. Inspection checks recognized paths, compatibility declarations, retained contents and native-record constraints. Review-scope validation checks its stored state, snapshot identity, ownership and associated inventory information. These safeguards address a changing workspace and a transferred snapshot at different stages of the exchange.

Import assigns receiving-workspace identities and remaps native references. Correspondence must therefore be evaluated through an identity map and the meanings of relations, rather than requiring every internal identifier or metadata checksum to remain unchanged. Historical tool-argument payloads can still contain source identifiers; these are interpreted through origin information rather than assumed to be rewritten local foreign keys. The source package and origin information remain available as the reference for the imported history. Recorded history is read-only; subsequent discussion or continuation uses an associated writable context and receiving-side permissions. Import does not automatically execute the retained code or restore source credentials.

Figure 1 summarizes the preserved relationships and permitted changes in this exchange contract.

\begin{figure}[!htbp]
\centering
\includegraphics[width=\linewidth,height=0.72\textheight,keepaspectratio]{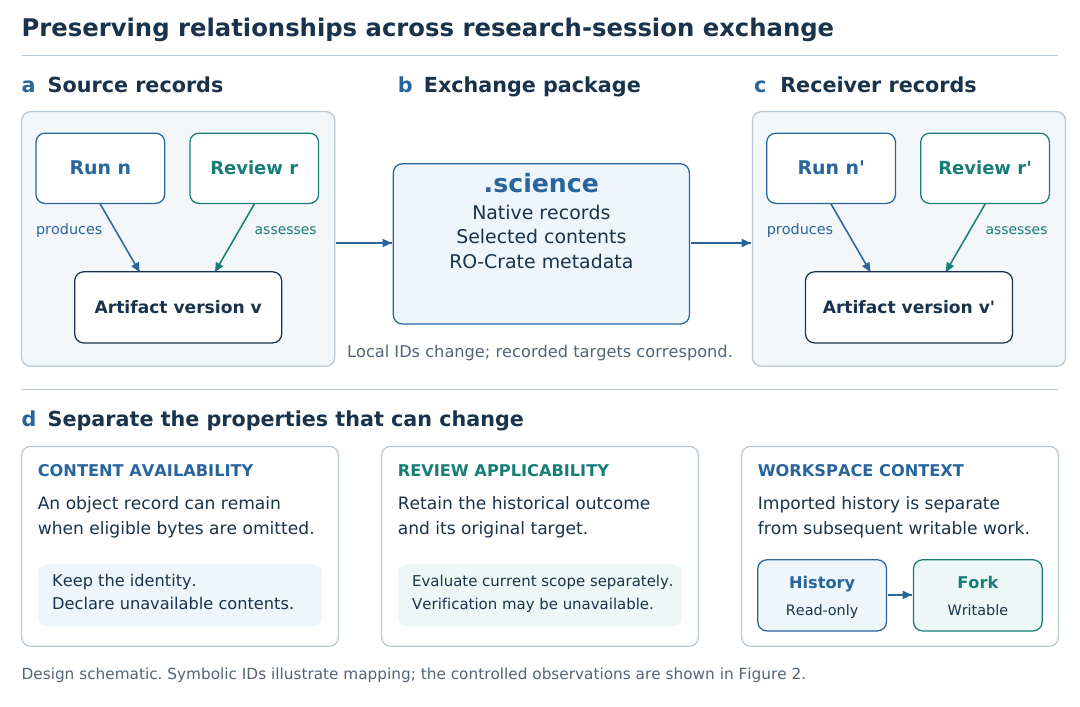}
\caption{Preserved relationships and permitted changes in research-session exchange. (a--c) A representative source subgraph links a producing run and an assessment to a particular artifact version. The \texttt{.science} package carries native records, selected contents and RO-Crate metadata. Receiver identifiers may differ, while their mapped references must retain the corresponding targets. Symbols n, r and v denote illustrative identifiers, not measured record counts; the diagram omits other native record types. (d) The contract distinguishes object identity from payload availability, historical assessment outcomes from current scope applicability, and imported history from a writable fork. This is a design schematic rather than an additional experiment; the controlled observations appear in Figure 2.}
\end{figure}
\subsection{Implementation freeze and supporting evidence}
Controlled operations used Open-Science 0.37.1 on macOS. The freeze separately records the inspected main-branch source, the release-tag source and the installed application; their exact identifiers and integrity records are provided in the supporting version table. Seven relevant modules were compared with the installed backend's source map. Six matched exactly; the native snapshot module differed in import paths and type-import organization. This checks correspondence for the examined mechanisms without attesting the entire application build.

Production-function probes used synthetic records at a separately pinned source revision to test artifact-version review projection, file-selection validation, retained-copy conflicts and selection preview. All nine assertions passed under Node.js 24.19.0. The two tested source files retained the hashes recorded in the implementation freeze. Individual assertions and outputs remain in the supporting materials; these function tests have a separate scope from the controlled application exchange.

The registry example uses a case16 package frozen on 12 October 2026 and retained unchanged in the accompanying materials. The supporting audit records provide its exact package identifier and integrity checksum. Inspection checked every inventory entry and followed selected task, execution and review references. The digest identifies the evaluated snapshot independently of subsequent changes to the public case collection.

We extracted the package's retained \texttt{analysis.py}, \texttt{registry\_\allowbreak{}source.json} and two reference CSV tables, then executed the inspected script offline with Python 3.12.14 and the standard library. The command was \texttt{python3 analysis.py registry\_\allowbreak{}source.json --out rerun}. Byte comparisons assessed the regenerated tables against their retained versions. This fixed-input calculation used saved registry data; it did not import the package into a fresh application or replay the original agent session.

\subsection{Controlled paired validation and the exchange contract}
The protocol was frozen before the controlled application run. Five points, x=1,2,3,4,5 and y=3,5,7,9,11, specified independently known values of n=5, slope=2, intercept=1 and mean=7. The task required regression and an explicitly planned but unexecuted permutation test. The first output intentionally contained mean=8 as a disclosed fixture. A manual Reviewer request produced a failed finding, then automatically initiated correction and rechecking. The corrected version contained mean=7. This differed from the anticipated manually staged correction. An operation addendum preceded a further revision carrying an unreviewed fixture marker, without another review request. Automatic correction, failed input-access attempts and attachment-only messages remain in the history.

Source snapshots extracted independently of the exporter combined read-only, session-scoped database queries, persisted conversation and owned Notebook files. Checkpoints captured the first output before correction, the reviewed correction and the final source before export. These snapshots provide an exporter-independent reference for recorded state; they do not establish that the application captured every relation in the underlying research process. Complete and partial-content packages were generated through the application export interface. The partial selection excluded an unused uploaded note. An attempted exclusion of the computational CSV was prevented by its required-evidence control.

Both packages were imported through the project menu's \textbf{Import Session package} entry into separate, initially empty projects within the same installed application. The auditor selected explicit native fields for object identity, revision direction, producer and prompt references, review ownership and scope, findings, dispositions, uploads and message snapshots. Source-to-export checks compare the saved values directly. Source-to-receiver checks map source expectations through the saved import receipt and compare them with the actual receiver values; they never apply the identity map to receiver observations. The auditor parses structured JSON fields for value comparison, treats absent optional values as null and requires each compared object to exist. A missing referenced object or a mismatched expected target cannot pass as an equal relation. The full comparison lists identify the selected fields and their values; they do not claim exhaustive reference validation.

For each of two review snapshots, receiver checks compare project, session, review, snapshot identity and scope. They allow rewritten file paths and serialization to differ. The auditor does not reuse the production identity-remapping function. The auditor also checks the mapped identity, role and content of each retained message, including the three nonempty user prompts. These checks cover the named message fields, not every nested tool-activity field. Two deliberately altered copies test the auditors themselves: an unremapped receiver producer reference and an inconsistent export-reference version number must produce failing checks and nonzero exit statuses.

Codex assisted the application operations and saved-record audits under researcher direction; no independent recipient participated. The operator inspected the three summary-version review views and used \textbf{Fork to continue} to create a distinct writable context. The operation report records comparisons of the source and read-only imported session files with their frozen snapshots after the fork. The retained materials include that report and the fork snapshot, but not a complete UI recording or separate post-fork copies of both files. Offline audits therefore recheck the saved records without repeating those application observations.

Table 1 concentrates the export-and-import contract. Correctness requires the appropriate object, version, direction and assessment scope; local identifiers may change. A historical failed finding can remain correctly preserved while a newer version remains unreviewed. A retained instruction and an identifiable omitted object remain distinct from executed activity and available content, respectively.

{\fontsize{8}{10}\selectfont\setlength{\tabcolsep}{3pt}
\begin{longtable}{>{\raggedright\arraybackslash}p{1.200in}>{\raggedright\arraybackslash}p{1.200in}>{\raggedright\arraybackslash}p{1.200in}>{\raggedright\arraybackslash}p{1.200in}>{\raggedright\arraybackslash}p{1.200in}}
\multicolumn{5}{l}{\fontsize{11}{13}\selectfont \textbf{Table 1. Export-and-import contract for the evaluated research-session records.}} \\[6pt]
\toprule
Record or condition & Native fields or retained evidence & Required meaning after exchange & Permitted change & Omission or refusal rule \\
\midrule\endhead
Task context and producing activity & Conversation; \texttt{prompt\allowbreak{}Message\allowbreak{}Id}, \texttt{producer\allowbreak{}Run\allowbreak{}Id}, message snapshots; Notebook activity & The referenced turn and producing activity remain locatable; surrounding instruction is not invented as a direct edge & Message and run identities may be remapped & Missing required referenced evidence is not silently replaced by a narrative \\
\addlinespace
Planned versus executed work & Task and protocol text; retained code, outputs and run status & A described plan does not become an execution event & Local execution context may differ & Absence of retained execution is reported within the inspected trajectory \\
\addlinespace
Input identity and access & Uploaded version and contents; available input associations; Notebook access evidence & Identity, evidence strength and content availability remain distinguishable & Version identity and storage path may be remapped & Explicit content exclusions do not imply available bytes or a stronger dependency claim \\
\addlinespace
Artifact revision & Lineage, \texttt{version\allowbreak{}Number}, \texttt{based\allowbreak{}On\allowbreak{}Version\allowbreak{}Id}, \texttt{current\allowbreak{}Version\allowbreak{}Id} & Distinct versions and recorded revision direction remain identifiable & Artifact and version identifiers may be remapped & Required referenced versions are captured with their records \\
\addlinespace
Finding and corrective assessment & \texttt{artifact\allowbreak{}Version\allowbreak{}Id}, review \texttt{scope}, scope snapshot, finding disposition and \texttt{assessed\allowbreak{}Artifact\allowbreak{}Version\allowbreak{}Id} & Historical and corrective judgments retain their respective targets; no pass is assigned to an unreviewed revision & Record identifiers may be remapped; current applicability requires a separate scope comparison & Required scope evidence is validated; a historical judgment is not reissued as a new check \\
\addlinespace
Selected contents & Inventory and explicit exclusion descriptors & An omitted payload remains distinguishable from available content & Receiver storage paths may change & Only eligible exclusions are accepted; inseparable retained-copy conflicts are guarded \\
\addlinespace
Imported history and new work & Origin information, import receipt, historical session and fork & Preserved history remains separate from writable continuation & Receiving project and session identifiers change & Source permissions and active computation are not restored automatically \\
\addlinespace
\bottomrule\end{longtable}}
The information-location comparison used a documentation-rich ordinary directory from the same source snapshot. It retained eight artifact and upload versions, original conversation, Notebook evidence, original review records and a file identity index. Seven predefined questions examined task context, completed activity, the unexecuted plan, input access, version identity, assessment scope and full-export content availability. An unblinded operator/script audit recorded whether supporting evidence could be located in each representation. Original review bindings remained in the control. Separation of historical and writable contexts was evaluated through application import and fork, outside the static comparison.

\section{Results}
\subsection{The implemented package carries relational records as well as files}
Source inspection identified native fields for artifact derivation, input association, task and producing-run references, finding-to-version binding, finding dispositions and review-scope snapshots. The inspected exporter implements traversal over these relation types; the controlled trajectory exercises the populated version, review and snapshot relations described below. It contained no native \texttt{Artifact\allowbreak{}Version\allowbreak{}Input} rows. File selection acts on eligible payloads under the record-retention rules, and import code constructs receiving-workspace mappings for native records and conversation objects. This implementation evidence describes supported mechanisms separately from the relations exercised in the controlled session.

The distinction between records and contents was visible in the production code. Known exclusions were validated against stored metadata, and retained inventory entries with the excluded content's digest triggered an error. These rules define two responses to partial sharing. Eligible contents can be omitted with descriptors; conflicting selections are refused when the bytes remain inseparable from retained evidence. Review projection independently selected findings applicable to a chosen version, preserving historical review events without replacing that version's direct assessment.

\subsection{The controlled source separated revision and assessment events}
The final source contained three artifact lineages and six artifact versions. Additional records comprised two uploaded versions, two reviews, two findings, two review-scope snapshots and one finding disposition. The three summary versions represented the injected error, its reviewed correction and a further unreviewed revision. Their producing Notebook activities and version references were separately recorded. The failed finding remained bound to the first version. Its disposition linked resolution to assessment of the corrected version. No matching scope assessed the final revision.

Task interpretation required the surrounding conversation. The first version's prompt reference identified the later CSV attachment-only turn; the initial instruction remained in an earlier message. Both records were preserved without replacing the attachment turn with a reconstructed instruction. The protocol described the permutation test as unexecuted. Retained Notebook cells documented regression, correction and fixture writing, but no implementation of that test. Input use was traceable through retained CSV contents and the producing Notebook code.

Figure 2 summarizes the controlled trajectory and the two exchange conditions.

\begin{figure}[!htbp]
\centering
\includegraphics[width=\linewidth,height=0.72\textheight,keepaspectratio]{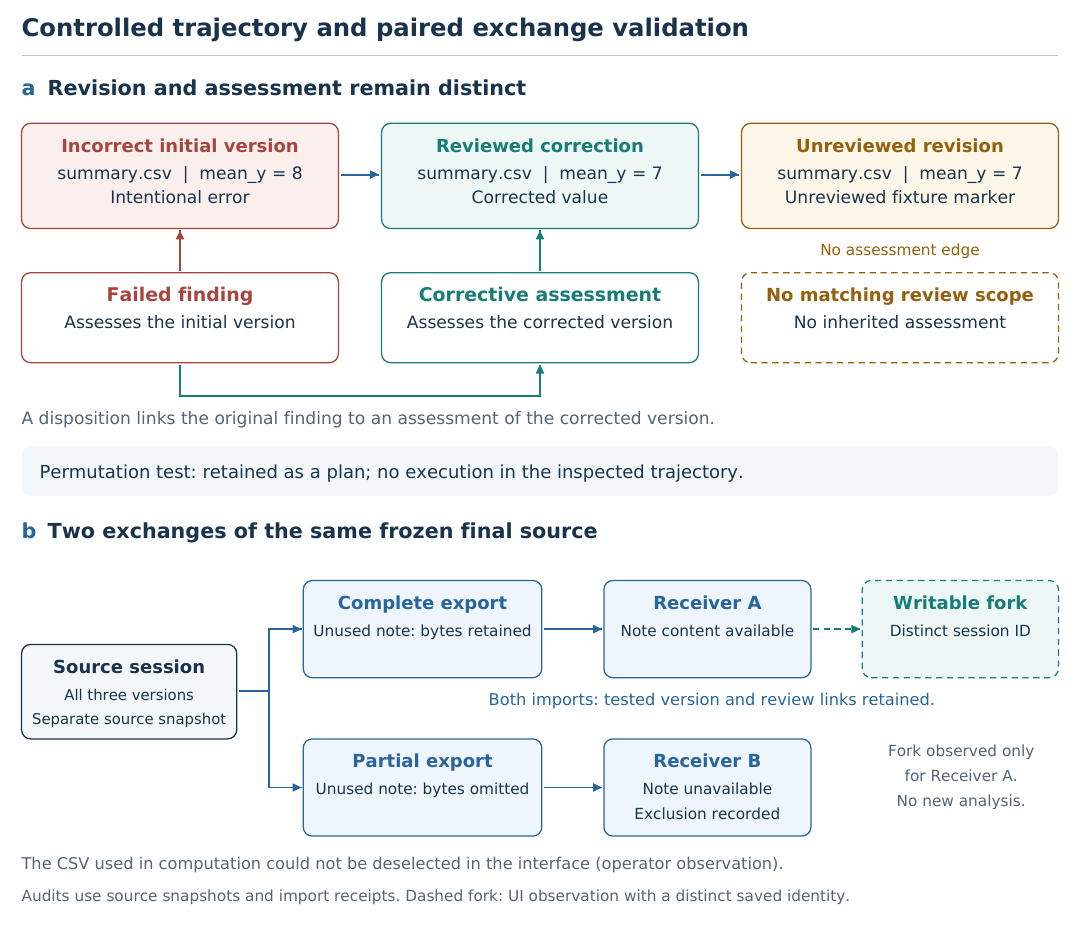}
\caption{Controlled revision trajectory and paired exchange conditions. (a) Three versions of \texttt{summary.csv} retain distinct assessment states. The initial version contains the intentional mean error; the corrected version has a corrective assessment; the subsequent revision carries an unreviewed fixture marker without a matching review scope. The disposition links the original finding to the corrective assessment without assigning that assessment to the unreviewed revision. The permutation test remains a recorded plan within the inspected trajectory. (b) Complete and partial exports originate from the same source session, with independently extracted snapshots serving as the audit reference. Both receiving projects retain the tested version and review associations; only the partial condition omits the unused note's bytes and records that exclusion. Both imports use the same application installation. The dashed fork marks the operator's editable-context observation for the complete import and the distinct saved session identity; it does not represent a new scientific analysis. The computational CSV's required-evidence safeguard was an interface observation.}
\end{figure}
\subsection{A public registry case connects a table with task, execution and review}
The evaluated case16 package contained ten artifact versions, two retained message snapshots, two reviews, eight findings and two review-scope snapshots. All 100 inventory entries matched their listed sizes and SHA-256 digests. The audit identifies this copy by its measured size of 1,846,866 bytes and the package integrity record in the supporting materials.

For \texttt{trial\_\allowbreak{}calculations.csv}, the native artifact-version record referenced a task message present in the session, a producing Notebook run present in the retained Notebook document and an available message snapshot. The producing run had a completed status. A finding was bound to the same artifact version with a scope-validated binding, and the parent review's scope contained that version. Its review-scope snapshot was also retained. RO-Crate metadata connected a retained file object to the native version identity. These references allowed a concrete query to be answered: which task, recorded computation and scoped check were associated with the exported table?

The session addressed registry-results availability in a bounded glioblastoma cohort. Re-executing the retained calculation considered 93 candidate records and excluded 13 under its screening rule, leaving 80 included records. These counts describe the example analysis rather than agent performance. The offline run regenerated \texttt{trial\_\allowbreak{}calculations.csv} and \texttt{availability\_\allowbreak{}summary.csv} byte for byte from the saved script and normalized source. Script, input and output hashes, runtime version and execution output accompany the package.

The package contained zero native \texttt{Artifact\allowbreak{}Version\allowbreak{}Input} rows. Input information was nevertheless locatable in the retained source, script and execution evidence, supporting inspection of observed use without establishing an exhaustive dependency graph. The case supplies a research example and a verified fixed-input reuse path. It has no independently captured original-session reference for a paired source-to-export measurement. Figure S1 summarizes the retained associations and the fixed-input re-execution in the Supplementary Information.

\subsection{Paired source-to-import validation}
Complete and partial-content exports each matched the exporter-independent source reference on all 138 specified native record-field comparisons. Each receiver matched on 136 mapped record-field comparisons and ten snapshot-semantic comparisons (Table 2). The latter replace the two whole snapshot-serialization checks used at export: each of two snapshots receives five semantic checks. Both receivers contained 28 messages; each mapped message identity was unique and its role and content matched, including all three nonempty user prompts. The complete and partial exports also passed 52 and 50 inventory size-and-digest checks, respectively.

{\fontsize{8}{10}\selectfont\setlength{\tabcolsep}{3pt}
\begin{longtable}{>{\raggedright\arraybackslash}p{0.983in}>{\raggedright\arraybackslash}p{0.983in}>{\raggedright\arraybackslash}p{0.983in}>{\raggedright\arraybackslash}p{0.983in}>{\raggedright\arraybackslash}p{0.983in}>{\raggedright\arraybackslash}p{0.983in}}
\multicolumn{6}{l}{\fontsize{11}{13}\selectfont \textbf{Table 2. Fixed audit units for the two exchange conditions.}} \\[6pt]
\toprule
Condition and stage & Other native record fields & Whole snapshot serialization & Snapshot semantic fields & Total specified comparisons & Matched \\
\midrule\endhead
Complete export & 136 & 2 & --- & 138 & 138 \\
\addlinespace
Partial-content export & 136 & 2 & --- & 138 & 138 \\
\addlinespace
Complete import & 136 & --- & 10 & 146 & 146 \\
\addlinespace
Partial-content import & 136 & --- & 10 & 146 & 146 \\
\addlinespace
\bottomrule\end{longtable}}
The counts are fixed comparisons within one source trajectory and related exchange conditions, not independent samples or estimates of capture completeness. Message checks and inventory checks are separate from the table totals. The accompanying CSV files list the source, expected mapped and observed values. Both deliberately altered audit copies failed as required: the receiver producer-reference mismatch appeared as a failed field comparison, and both export conditions detected the changed source version number. These negative checks validate the offline auditors, not application behavior under injected failures.

In the application inspection, the operator observed the failed finding on the first summary version, the tracked correction on the second and \textbf{No review for this version} on the final revision. The saved records independently retained these version and scope associations. The operator also reported outdated labels on historical conversation review cards. The retained operation record does not establish whether those labels already existed in the final source view or which scope difference triggered them after import; §5.3 separates this unresolved presentation detail from the verified historical bindings.

Selective export retained the unused note's version record in both conditions. Its content was accessible after complete import but absent after partial import, with an explicit exclusion descriptor. The interface marked the computational CSV as \textbf{Required evidence}, preventing its deselection and showing its Notebook use. This established an interface-level selection safeguard; the actual session test did not submit a contradictory request to the backend, whose refusal mechanism was instead covered by the supplementary function probes.

The operator observed read-only imported history with execution and continuation disabled. \textbf{Fork to continue} produced a distinct session with an editable composer, and the saved fork had a different session identity. The recorded post-fork file comparison reported that the source and imported history remained unchanged. No new analysis was requested in the fork, so the observation concerns context separation. Table 3 summarizes the relationship-level results and distinguishes saved-record checks from application observations.

{\fontsize{8}{10}\selectfont\setlength{\tabcolsep}{3pt}
\begin{longtable}{>{\raggedright\arraybackslash}p{1.525in}>{\raggedright\arraybackslash}p{1.525in}>{\raggedright\arraybackslash}p{1.525in}>{\raggedright\arraybackslash}p{1.525in}}
\multicolumn{4}{l}{\fontsize{11}{13}\selectfont \textbf{Table 3. Observed preservation and interpretation in the controlled exchange.}} \\[6pt]
\toprule
Question & Source reference or observation & Export and receiving-side observation & Interpretation \\
\midrule\endhead
Task and producing activity & Initial instruction, attachment turn and completed producer & Messages and recorded producer references remained locatable after mapping & Retained context and activity association; initial task attribution remains contextual \\
\addlinespace
Unexecuted step & Prespecified permutation plan and inspected run trajectory & Plan text retained; no new execution event introduced & Plan and observed execution remain distinguishable without claiming a formal Plan graph \\
\addlinespace
Input access & Supplied CSV and actual Notebook access & Input content and relevant code retained; CSV shown as required evidence & Supports inspection of observed use, not exhaustive dependency capture \\
\addlinespace
Artifact revisions & Incorrect first output, reviewed correction and unreviewed revision & Distinct contents, version direction and producer references preserved & Three revisions remained distinguishable \\
\addlinespace
Version-specific checks & Failed finding, corrective disposition and absent final-revision scope & First-version failure, tracked correction and final-revision no-review presentation & Historical checks were not reassigned to the unreviewed revision \\
\addlinespace
Partial-content availability & Unused note permitted for exclusion & Identity retained, exclusion declared and bytes unavailable only in partial import & Identity and payload availability remained separate \\
\addlinespace
Selection safeguard & CSV associated with retained execution evidence & Required-evidence control prevented deselection & Interface prevention observed; no end-to-end backend rejection claimed \\
\addlinespace
Historical versus writable context & Saved imported record and fork; operator report & Distinct fork identity; operator observed read-only history and editable fork, and reported source/history unchanged & Context separation; no new analysis or exhaustive write-prevention test \\
\addlinespace
\bottomrule\end{longtable}}
\subsection{Static evidence availability across two delivery representations}
All seven predefined static inspection questions had locatable supporting evidence in both the complete \texttt{.science} export and the documentation-rich ordinary directory (7/7 in each representation). The ordinary condition retained original conversation, Notebook and review metadata with all file versions and a locator index, including version-specific review bindings. Task attribution and unexecuted-plan judgments required the same contextual interpretation in both representations.

This comparison shows that a carefully documented conventional delivery can convey the inspected facts. The controlled application exchange separately tested native associations, explicit content availability and imported-history behavior; the ordinary directory did not undergo a corresponding application import or fork. The seven static checks were an unblinded operator/script audit, with no independent recipient answers or recovery-time measurement.

\section{Discussion}
\subsection{Design lessons from versioned research-session exchange}
The controlled trajectory exposed a distinction that a final output alone cannot express: an earlier failed assessment, a corrective assessment and the absence of a check on a new revision can all be true at once. Their exchange requires preserving the targets of historical judgments while evaluating each requested version separately. In this implementation, native version references, review scopes and finding dispositions provide that separation. The checked receiver records retained the relations, and the operator observed the final revision without an inherited assessment.

Partial sharing exposed a second distinction. The unused note retained an identifiable version record even when its bytes were absent. The computational CSV, by contrast, was protected by the interface's required-evidence control. The resulting design rule is conditional: represent eligible omissions explicitly, and refuse selections that conflict with inseparable retained evidence. The application observation and the function probes support different parts of that rule. They do not demonstrate how every downstream analysis or review would behave with a missing computational input.

These observations also delimit the record's explanatory power. The producing version referenced an attachment-only turn, so attribution to the earlier research instruction still required reading the conversation. Notebook code documented input access despite the absence of native input-edge rows. Preserving such records allows a recipient to inspect their meaning; it cannot turn a contextual association into an explicit causal dependency.

\subsection{Contribution within existing provenance and agent infrastructure}
AiiDA supplies consistency-aware export, provenance abstraction studies partial disclosure, and RO-Crate provides research-object descriptions [7--10,19]. OpenAI4S integrates scientific session exchange, while agent-provenance and artifact-observability systems organize interactions, evidence and evaluator bindings [11--15]. The design contribution here is an implemented contract for how workbench-managed version and review relations survive local identity mapping and selected content omissions. Its validation contribution is the paired source/export/receiver examination of those rules under a controlled sequence of revision and assessment events.

The ordinary-directory result helps locate that contribution: adequate documentation supplied locatable evidence for all seven questions. The application exchange test asks a different question, whether selected records retain their native associations and explicit availability while providing a separate writable context. These results support that behavior in the tested installation. They offer concrete implementation choices and checks for session exchange, without measuring comparative user understanding or efficiency.

\subsection{Applicability, capture limits and observation boundaries}
Historical outcome and current applicability answer different questions. A failed finding records a judgment about its original target; a resolved disposition records a later corrective assessment. Neither assigns a result to an unreviewed revision. The inspected implementation separately recomputes a completed review's scope and compares relevant frame/branch identities and the sequence of block source identifiers and content hashes. A difference marks the review stale; inability to recompute sets verification unavailable rather than establishing that the content changed.

This mechanism explains why a historical outcome can remain intact while its current applicability needs reassessment. It does not by itself explain the outdated labels observed in this experiment. The saved evidence lacks the corresponding before-and-after UI states and resolved-scope differences, so their specific trigger remains undetermined. We verified the specified historical bindings and retained the operator's presentation observation as a separate result. Allowing derived applicability to change does not make every possible display difference correct.

The main exchange test used one controlled trajectory and two receiving projects in the same installation. Both the controlled source and case16 had no native \texttt{Artifact\allowbreak{}Version\allowbreak{}Input} rows, although their retained code and data documented observed input use. Source capture completeness therefore remains unmeasured. The case16 rerun adds evidence for two fixed-input outputs, not fresh data acquisition or replay of the complete agent process. The fork observation established a writable context without testing scientific continuation; its post-operation immutability comparison remains an operator report because the delivery lacks separate post-fork copies of both compared files.

Missing-evidence refusal, concurrent-source-change safeguards and scope validation were identified in code. The paired application test did not inject failures for every guard. The nine function probes and two offline-auditor negative checks accordingly have narrower roles than end-to-end application validation. A future extension could test these failure paths, cross-installation exchange and recipient interpretation using the same explicit separation of preserved records, missing contents and current applicability.

\subsection{Future research handoff}
A planned human-readable research account would add an evidence-linked reading layer to the current structured records. It would explain task context, executed and unexecuted steps, inputs, revisions, assessment scope, known omissions and continuation conditions. Substantive statements should link to their supporting messages, activities, versions or snapshots, while distinguishing recorded facts from interpretation. The separately prepared Hugging Face READMEs remain public explanations outside the export workflow, not evidence that this planned account already exists. Future evaluation could examine evidence fidelity and recipient understanding. This extension would complement portable records as infrastructure for research sharing, publication-linked inspection and continued collaboration.

\section{Conclusion}
AIPOCH Open-Science implements a research-session exchange contract that preserves recorded version and assessment relationships while permitting local identifier mapping and explicit omission of eligible contents. A controlled same-installation trajectory retained the tested relations through complete and partial exchange, distinguished an unreviewed revision from earlier checks and produced a separate writable fork. The case16 inspection linked a scientific table to its recorded task, execution and review, and a fixed-input run reproduced two saved tables byte for byte. Together, these results provide design rules and a concrete validation procedure for exchanging recorded research work. Their scope remains the tested records and operations; broader capture completeness and recipient outcomes require further evaluation.

\section*{Data and code availability}
AIPOCH Open-Science is available under the Apache License 2.0 at \url{https://github.com/aipoch/open-science}; the official website is \url{https://aipoch.com}. The public research-case collection is \url{https://huggingface.co/datasets/AIPOCH-AI/open-science-cases}. The evaluated case16 snapshot is identified in the validation release by separate integrity records for the original export and its de-identified publication copy. Its identity is not inferred from the case collection's current main branch.

The validation materials are publicly available in a \href{https://huggingface.co/datasets/AIPOCH-AI/case-validation/tree/5a775981cc7142651116960b0b96501088670c98/releases/2026-10-13-deidentified}{versioned, de-identified release at AIPOCH-AI/case-validation on Hugging Face}. They include the protocol and amendments, exporter-independent source snapshots, complete and partial packages, import receipts, per-field comparisons, the ordinary-directory control and the operator's fork report. Offline auditors, negative checks, pinned function probes, the case16 publication package and its fixed-input rerun are also included. Technical identifiers are consistently aliased and absolute local paths are replaced by stable opaque references. Package inventories and RO-Crate file-integrity metadata describe the publication copies; recorded historical snapshot fingerprints retain their original meaning. Application observations concern the original records, whereas renewed offline audits check the publication copies. The release README maps files to manuscript results and documents runtimes, commands, access and reuse conditions, and evidence boundaries. A transformation record and file manifest distinguish the public copies from the original evidence, which remains preserved locally.

\section*{Author contributions}
AIPOCH contributors designed and developed the product, implemented the research infrastructure and conducted associated development work. Collaborating institutional authors participated in use, supplied feedback and revision suggestions, and supported supplementary product validation.

\section*{Declaration of AI assistance}
Codex (OpenAI) assisted with literature retrieval, source-code inspection, application operation, evidence organization, manuscript revision, validation code and offline re-execution. The Methods and supporting materials identify the scope of these activities. The authors are responsible for checking sources, interpreting findings and reviewing the submitted text and materials.

\clearpage\section*{References}
\begingroup\fontsize{9}{10}\selectfont\raggedright
\begin{enumerate}\setlength{\itemsep}{2pt}\setlength{\parsep}{0pt}\setlength{\parskip}{0pt}
\item Alizadeh, M., Mosleh, M., Gilardi, F., Kasirzadeh, A., \& Tucker, J. (2026). \emph{AI coding agents can reproduce social science findings} [Preprint]. arXiv. \url{https://arxiv.org/abs/2606.11447}
\item Alizadeh, M., Gilardi, F., Mosleh, M., \& Kasneci, E. (2026). \emph{AI coding agents in social science: Methodologically diverse, empirically consistent, interpretively vulnerable} [Preprint]. arXiv. \url{https://arxiv.org/abs/2606.11456}
\item Dunn, C. W., Schultz, D. T., \& Musser, J. M. (2026). Designing reproducible large-language-model-assisted scientific analyses. \emph{Patterns, 7}(9), 101644. \url{https://doi.org/10.1016/j.patter.2026.101644}
\item Samuel, S., \& Mietchen, D. (2024). Computational reproducibility of Jupyter notebooks from biomedical publications. \emph{GigaScience, 13}, giad113. \url{https://doi.org/10.1093/gigascience/giad113}
\item Lebo, T., Sahoo, S., \& McGuinness, D. (Eds.). (2013). \emph{PROV-O: The PROV ontology}. W3C Recommendation. \url{https://www.w3.org/TR/prov-o/}
\item Garijo, D., \& Gil, Y. (2013). \emph{The P-Plan ontology} (Revision 1.1). \url{https://www.opmw.org/model/p-plan/}
\item Soiland-Reyes, S., Sefton, P., Crosas, M., Castro, L. J., Coppens, F., Fernández, J. M., Garijo, D., Grüning, B., La Rosa, M., Leo, S., Ó Carragáin, E., Portier, M., Trisovic, A., RO-Crate Community, Groth, P., \& Goble, C. (2022). Packaging research artefacts with RO-Crate. \emph{Data Science, 5}(2), 97--138. \url{https://doi.org/10.3233/DS-210053}
\item Leo, S., Crusoe, M. R., Rodríguez-Navas, L., Sirvent, R., Kanitz, A., De Geest, P., Wittner, R., Pireddu, L., Garijo, D., Fernández, J. M., Colonnelli, I., Gallo, M., Ohta, T., Suetake, H., Capella-Gutierrez, S., de Wit, R., Kinoshita, B. P., \& Soiland-Reyes, S. (2024). Recording provenance of workflow runs with RO-Crate. \emph{PLOS ONE, 19}(9), e0309210. \url{https://doi.org/10.1371/journal.pone.0309210}
\item Huber, S. P., Zoupanos, S., Uhrin, M., Talirz, L., Kahle, L., Häuselmann, R., Gresch, D., Müller, T., Yakutovich, A. V., Andersen, C. W., Ramirez, F. F., Adorf, C. S., Gargiulo, F., Kumbhar, S., Passaro, E., Johnston, C., Merkys, A., Cepellotti, A., Mounet, N., \ldots{} Pizzi, G. (2020). AiiDA 1.0, a scalable computational infrastructure for automated reproducible workflows and data provenance. \emph{Scientific Data, 7}, 300. \url{https://doi.org/10.1038/s41597-020-00638-4}
\item AiiDA developers. (n.d.). \emph{Consistency}. AiiDA 2.9.3 documentation. Retrieved October 12, 2026, from \url{https://aiida.readthedocs.io/projects/aiida-core/en/stable/topics/provenance/consistency.html}
\item Zhang, G., Li, H., Wang, Y., Lin, M., Lv, L., Mao, Y., Wang, Y., Zhu, J., Tang, M., Jiang, Z., Wang, Y., Yao, J., Ning, K., Pang, D., Tian, Y., OpenAI4S Community, Liu, Y., \& Yuan, L. (2026). \emph{OpenAI4S: Code as action, science as sessions} (Version 1) [Preprint]. arXiv. \url{https://arxiv.org/abs/2609.15096v1}
\item Souza, R., Gueroudji, A., DeWitt, S., Rosendo, D., Ghosal, T., Ross, R., Balaprakash, P., \& Ferreira da Silva, R. (2025). \emph{PROV-AGENT: Unified provenance for tracking AI agent interactions in agentic workflows} [Preprint]. arXiv. \url{https://arxiv.org/abs/2508.02866}
\item Wu, E., Castelo, S., Liu, Y., Silva, C. T., \& Freire, J. (2026). \emph{AgentTrails: Towards trust and reuse for agentic tasks} [Preprint]. arXiv. \url{https://arxiv.org/abs/2607.18816}
\item Yin, X., Du, M., Prince, M. H., \& Cherukara, M. J. (2026). \emph{Artifact-centered claim-aware observability for autonomous scientific agents} [Preprint; accepted at the LM4Sci Workshop, COLM 2026]. arXiv. \url{https://arxiv.org/abs/2608.18312}
\item Krebs, F. (2026). \emph{F(AI)2R: Who did what, and who checked? Verifiable AI provenance as an executable skill} [Preprint]. arXiv. \url{https://arxiv.org/abs/2607.25637}
\item Stanford SNAP. (n.d.). \emph{Biomni} [Software and documentation]. GitHub. Retrieved October 12, 2026, from \url{https://github.com/snap-stanford/Biomni}
\item serena2z. (2025, September 12). \emph{Download research as doc or pdf} (Issue \#206). Biomni GitHub repository. \url{https://github.com/snap-stanford/Biomni/issues/206}
\item Li, O., Agarwal, V., Zhou, S., Gopinath, A., \& Kassis, T. (2025). \emph{K-Dense Analyst: Towards fully automated scientific analysis} (Version 2) [Preprint]. arXiv. \url{https://arxiv.org/abs/2508.07043v2}
\item Missier, P., Bryans, J., Gamble, C., \& Curcin, V. (2020). Abstracting PROV provenance graphs: A validity-preserving approach. \emph{Future Generation Computer Systems, 111}, 352--367. \url{https://doi.org/10.1016/j.future.2020.05.015}
\item Research Object community. (n.d.). \emph{Five Safes Crate}. Retrieved October 12, 2026, from \url{https://www.researchobject.org/ro-crate/5s-crate}
\item Liu, J., Pei, J., Huang, J., Si, C., Qu, A., Tang, X., Lu, R., Chen, L., Bai, X., Zheng, H., Chen, C., Chen, Z., Ye, H., Fu, Y., He, Z., Jin, Z., Zhang, Z., Sun, S., Harmon, M., \ldots{} Zhang, Z. (2026). \emph{The last human-written paper: Agent-native research artifacts} (Version 3) [Preprint]. arXiv. \url{https://arxiv.org/abs/2604.24658v3}
\end{enumerate}\endgroup
\clearpage
\section*{Supplementary Information}
\setcounter{figure}{0}
\renewcommand{\thefigure}{S\arabic{figure}}
\begin{figure}[!htbp]
\centering
\includegraphics[width=\linewidth,height=0.72\textheight,keepaspectratio]{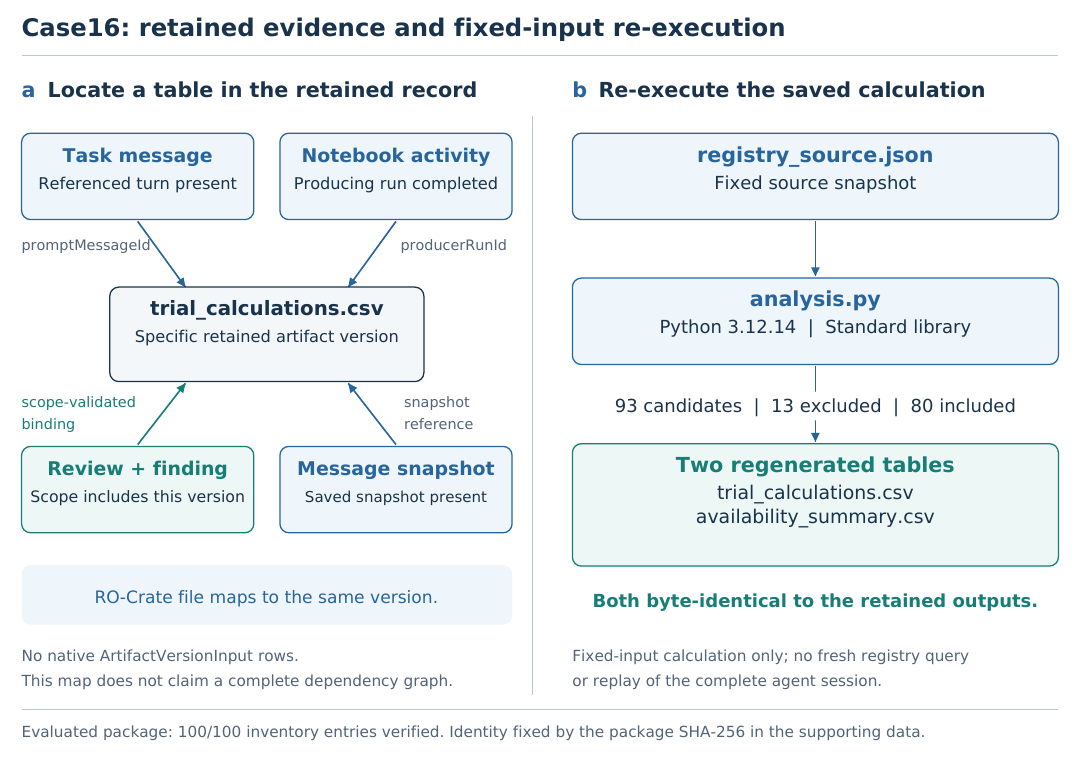}
\caption{Case16 evidence associations and fixed-input re-execution. (a) Retained references associate \texttt{trial\_\allowbreak{}calculations.csv} with a task message, a completed producing Notebook activity, a message snapshot and a scope-validated finding whose parent review includes the selected version. RO-Crate metadata also maps a file object to that version. Arrows gather these associations around the focal artifact; they do not assert an exhaustive causal graph. The package contains no native \texttt{Artifact\allowbreak{}Version\allowbreak{}Input} rows. (b) Offline execution of the retained \texttt{analysis.py} against \texttt{registry\_\allowbreak{}source.json} under Python 3.12.14 reproduces the two saved CSV tables byte for byte. The retained calculation considers 93 candidates, excludes 13 and includes 80. These numbers describe this fixed-source calculation, not a fresh registry query or a model-performance evaluation. All 100 inventory entries of the supplied package passed size and checksum checks.}
\end{figure}
\end{document}